\documentclass{ifacconf}

\usepackage{graphicx}
\usepackage{natbib}
\usepackage{amsmath}
\usepackage{amssymb}
\usepackage{xcolor}
\usepackage{bm}
\usepackage{enumitem}
\usepackage{booktabs}
\usepackage{multirow}

\usepackage{subcaption}

\usepackage{comment}

\begin{document}
\begin{frontmatter}

\title{Koopman-based NMPC for Virtually Coupled Train Control System
}


\author[First]{Yiwen Zhang}
\author[Second]{Lorenzo Calogero}
\author[First]{Shukai Li}
\author[Second]{Alessandro Rizzo}
\author[Second]{Anton V. Proskurnikov}

\address[First]{School of Systems Science, Beijing Jiaotong University, China\\
\{yiwenz, shkli\}@bjtu.edu.cn}
\address[Second]{Politecnico di Torino, Italy\\ \{lorenzo.calogero,alessandro.rizzo,anton.proskurnikov\}@polito.it}

\begin{abstract}  
This paper investigates an analytical Koopman-based nonlinear model predictive control (K-NMPC) approach for tracking control of virtually coupled train systems. A nonlinear train movement model incorporating train dynamics, speed and control input limits, passenger comfort constraints, and collision avoidance is systematically lifted into a finite-dimensional Koopman space through closed-form observable functions. After freezing the affine parameter-varying lifted predictor along the shifted predicted trajectory, the online optimal control problem is solved as a quadratic program that can be solved efficiently. The proposed K-NMPC is benchmarked against a time-discrete NMPC scheme, demonstrating comparable control performance with significantly reduced online computation time and strong potential for real-time implementation in practical virtually coupled train control systems.
\end{abstract}


\end{frontmatter}

\section{Introduction}
Virtually coupled trains, enabled by next-generation train signaling and control systems, provide a promising solution for achieving high-efficiency and high-density rail operations \citep{Mitchel:16}. By leveraging train-to-train (T2T) communication, virtually coupled trains operate through a coordination mechanism that allows multiple units to run cooperatively at short distances while maintaining consistent speeds \citep{Goikoetxea:16}. Virtually coupled train operation imposes stringent requirements on real-time information exchange, precise speed coordination, and strict safety assurance, posing significant challenges for the computational efficiency, control accuracy, and constraint-handling capabilities of control methods.

To reconcile the strict real-time requirements of virtually coupled train systems with high control performance, a common strategy combines local linearization of nonlinear train dynamics and safety constraints with MPC \citep{LiSK:15,LiuYF:22}. This approach enables the use of linear MPC, which provides advantages in handling state and input constraints while efficiently solving convex quadratic programs.
To better capture the intrinsic nonlinearities of traction and braking dynamics in virtually coupled trains, recent studies have explored nonlinear control strategies. \cite{Park:22} proposed a reference distance scheme and designed a robust distance controller based on sliding mode control. \cite{Su:22} developed a collision-avoidance controller using a Deep-Q-Network. \cite{ZhangQH:23} designed a backstepping adaptive controller to realize tracking control for virtually coupled trains. However, these approaches generally lack a unified framework to simultaneously enforce nonlinear dynamics and operational constraints, which may compromise performance in terms of smooth operation and efficient utilization of train resources.

For vehicle and train control systems with nonlinear dynamics and multiple operational constraints, nonlinear model predictive control (NMPC) has been widely applied due to its ability to explicitly incorporate nonlinear constraints within a unified optimization framework. NMPC has demonstrated notable success in tracking control~\citep{Felez:19,ZhangW:25} and trajectory planning~\citep{Boggio:23,Gao:25}. However, NMPC inherently involves nonconvex optimization, resulting in high computational costs and challenges in ensuring real-time feasibility and global optimality, particularly in safety-critical applications such as virtually coupled trains.

To improve computational efficiency and address modeling uncertainties, data-driven approaches such as neural network control, deep reinforcement learning, and fuzzy logic control have also been explored for nonlinear vehicle and train systems~\citep{Kuutti:21,Pliss:24,Faria:24,Santoso:18}. Although these methods can approximate complex control policies, their limited interpretability and difficulty in enforcing hard operational constraints remain problematic for safety-critical railway applications. Koopman-based nonlinear model predictive control (K-NMPC) offers an alternative route: by representing nonlinear dynamics in a lifted observable space, it enables nonlinear prediction models and constraints to be handled through linear or bilinear lifted representations~\citep{paper-goswami-2021,Korda:18,Calogero:25}. This makes it possible, after suitable finite-dimensional approximation and parameter freezing, to solve the online predictive control problem as a quadratic program with substantially reduced computational cost.



Koopman MPC has recently been applied to several train and vehicle control problems, including energy-efficient vehicle control, cooperative braking in urban rail, and robust vehicle motion control~\citep{WangP:25,GaoDZ:23,ZhengH:24}. Most existing approaches construct finite-dimensional Koopman models from data using EDMD or sparse identification. Such data-driven liftings depend on the training data and on the chosen dictionary of observables, often leading to high-dimensional or poorly interpretable predictors. This is particularly restrictive for virtually coupled trains, where collision-avoidance constraints contain physically meaningful nonlinear terms that must be enforced reliably in real time.


This work investigates a K-NMPC approach tailored to virtually coupled trains, in which both the nonlinear dynamics and nonlinear safety constraints are embedded analytically into the lifted observable space. By combining dimensionality reduction with constraint-preserving lifting, the proposed approach transforms the original nonconvex optimal control problem into a QP that can be solved efficiently in real time. As demonstrated in the simulation study, the proposed K-NMPC approach achieves significantly reduced computation times compared to a baseline NMPC scheme while maintaining comparable cooperative control performance for multiple virtually coupled trains.

The remainder of this paper is structured as follows. Section 2 formulates the optimal control problem for virtually coupled trains; Section 3 presents the standard NMPC framework as a baseline approach.
Section~4 develops the Koopman-based model predictive control approach. Section~5 presents numerical experiments to validate the proposed method. Finally, Section~6 concludes the paper.

\section{Problem Setup}

\begin{figure}[t]
\begin{center}
\includegraphics[width=0.99\linewidth]{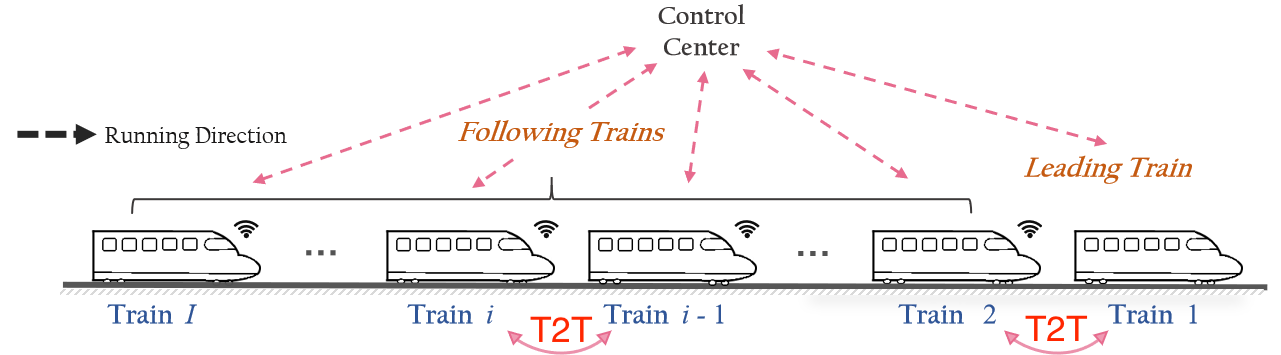}
\caption{Diagram of a virtually coupled train formation.}
\label{fig:VCTS}
\end{center}
\end{figure}

Consider a virtually coupled train formation consisting of $I \ge 2$ trains (Fig.~\ref{fig:VCTS}). Each train is indexed by $i \in {1,2,\ldots,I}$, where $i=1$ denotes the leading train and $i=2,3,\ldots, I$ represent the following trains.

\subsection{Train dynamics and constraints}

According to Newton's laws of motion, the train dynamics are as follows~\citep{Iwnicki:19,LiSK:20}:
\begin{eqnarray}\label{eq3-1}
\begin{cases}
\dot{p}_{i}(t) = v_{i}(t),\\
\dot{v}_{i}(t) = u_{i}(t)-(c_{0}+c_{1}v_{i}(t)+c_{2}v_{i}^2(t))-r_{i}(t),\\
\end{cases}
\end{eqnarray}
where $p_{i}(t)$ and $v_{i}(t)$ denote the position and speed of train $i$ at time $t$, respectively, and $u_{i}(t)$ is the control input representing the tractive force per unit mass when $u_{i}(t)>0$ or the braking force per unit mass when $u_{i}(t)<0$. The quadratic term $c_{0}+c_{1}v_{i}(t)+c_{2}v_{i}^2(t)$ represents the running resistance per unit mass, where $c_{0}$, $c_{1}$, and $c_{2}$ are resistance coefficients. The term $r_{i}(t)$ denotes the additional resistance per unit mass.

The state and control constraints are as follows:

(i) Speed limit constraints:
\begin{equation}\label{eq3-3}
0 \leq v_{i}(t) \leq v_{i}^{\max}(t),
\end{equation}
where $v_{i}^{\max}(t)$ is the maximum allowed speed at time $t$.

(ii) Control input (force) constraints:
\begin{equation}\label{eq3-4}
u_{i}^{\min} \leq u_{i}(t) \leq u_{i}^{\max},
\end{equation}
where $u_{i}^{\min}<0$ and $u_{i}^{\max}>0$ are the maximum braking and maximum tractive forces (per unit mass), respectively.

(iii) Jerk constraints:
\begin{equation}\label{eq3-5}
w^{\min} \leq \dot u_{i}(t) \leq w^{\max},
\end{equation}
where $w^{\min}<0$ and $w^{\max}>0$ are the lower and upper bounds on the rate of change of control input, respectively.

(iv) Collision avoidance constraints.
\begin{flalign}
& p_{i-1}(t) - p_i(t) - L_{i-1} \ge sm_0,  \label{eq3-6} \\
& p_{i-1}(t) - p_i(t) - L_{i-1}\ge sm_0 +
    \label{eq3-7}\\
& +\left(\frac{v_i(t)^2}{2b_i} + T_r v_i(t)-\frac{v_{i-1}(t)^2}{2b_{i-1}} \right), \nonumber
\end{flalign}
where $sm_0$ denotes the minimum safety distance, $L_{i-1}$ is the length of train $i-1$, $b_{i}$ is the emergency braking rate for train $i$, and $T_r$ is the reaction time delay.

\subsection{Objective function}
Different control objectives are implemented for the leading and following trains. The leading train operates as a guide for the virtually coupled train formation. We adopt a control policy for the leading train that tracks a reference trajectory while minimizing energy consumption. Therefore, the objective function for the leading train over the prediction horizon is formulated as:
\begin{equation} \label{eq3-8}
\begin{array}{ll}
{\rm min}~J_{i} & = \int_{t_{0}}^{t_{f}} \left(\zeta_{p,i}\left(p_{i}(t)-p_{r}(t)\right)^{2}\right. + \\
& \left.\zeta_{v,i}\left(v_{i}(t)-v_{r}(t)\right)^{2}+\zeta_{u,i}u_{i}^2(t)\right)dt, ~~i = 1,
\end{array}
\end{equation}
where $p_{r}(t)$ and $v_{r}(t)$ denote the reference position and speed profiles calculated by an external planner, $\zeta_{p,1}$ and $\zeta_{v,1}$ are weight coefficients for position and speed tracking errors, $\zeta_{u,1}$ is the weight coefficient for control effort, and $[t_{0},t_{f}]$ is the prediction horizon.

Since the virtually coupled train formation aims to operate as a convoy at high speed with short inter-train distances to improve operational efficiency, each following train adopts a control policy that tracks its preceding train to maintain the desired spacing and speed consistency. The objective function for each following train over the prediction horizon is formulated as:
\begin{equation} \label{eq3-9}
\begin{array}{ll}
{\rm min}~J_{i}  = &\int_{t_{0}}^{t_{f}} \left(\zeta_{p,i}\left(p_{i-1}(t) - p_{i}(t)-l\right)^{2}+ \right. \\
&+ \left.\zeta_{v,i}\left(v_{i-1}(t)-v_{i}(t)\right)^{2}+
\zeta_{u,i}u_{i}^2(t)\right)dt,\\&i = 2,\ldots, I,
\end{array}
\end{equation}
where $l = L_{i-1}+ d_{\text{des}}$ with $d_{\text{des}}$ denoting the desired spacing between consecutive trains, $\zeta_{p, i}$ is the weight for spacing tracking error, and $\zeta_{v, i}$ is the weight for speed tracking error relative to the preceding train.

\section{Baseline NMPC Approach}

In this section, a baseline NMPC approach is formulated for virtually coupled trains. The controller operates in discrete time, using the original nonlinear prediction model and constraints in the state space. Following the receding-horizon principle, a discretized finite-horizon optimal control problem is solved at each sampling instant. The first element of the optimal control sequence is then applied to the trains, and the procedure is repeated at the next sampling instant based on updated state feedback.


Under the MPC framework, a sampling period $T_s > 0$ is introduced for time discretization. Let $t_k = t_0 + kT_s$ denote the $k$-th sampling instant, with $k = 0,1,\ldots$. Consider the prediction horizon $[t_{k}, t_{k}+N_{p}T_{s}]$ starting from $t_k$, where $N_{p}$ is the prediction horizon length. The prediction step index $h = 0, \ldots, N_{p}$ corresponds to time instant $t_k + hT_s$. The notation $p_i(h,t_k)$, $v_i(h,t_k)$, and $u_i(h,t_k)$ denotes the predicted position, speed, and control input of train $i$ at prediction step $h$ at sampling instant $t_{k}$.

At each sampling instant $t_{k}$, the train dynamics~\eqref{eq3-1}, individual train constraints~\eqref{eq3-3}--\eqref{eq3-5}, and collision avoidance constraints~\eqref{eq3-6}--\eqref{eq3-7} can be rewritten as:
\begin{equation}\label{eqDmpc-1}
\begin{aligned}
p_{i}&(h+1,t_{k}) = p_i(h,t_k)+T_sv_{i}(h,t_{k}),\\
v_{i}&(h+1,t_{k}) = v_i(h,t_k)+T_s\left(u_{i}(h,t_{k})\right.\\
&-(c_{0}+c_{1}v_{i}(h,t_{k})+c_{2}v_{i}(h,t_{k})^2)-\\
&\left.-r_i(h,t_k)\right),\quad h = 0,\ldots,N_{p},
\end{aligned}
\end{equation}
\begin{equation}
0 \leq v_{i}(h,t_{k}) \leq v_{i}^{max}(h,t_{k}), h = 1,\ldots,N_{p}+1,\label{eqDmpc-2}
\end{equation}
\begin{equation}
u_{i}^{\min} \leq u_{i}(h,t_{k}) \leq u_{i}^{\max},~~ h = 0,\ldots,N_{p}, \label{eqDmpc-3}
\end{equation}
\begin{equation}
\begin{array}{ll}
w^{\min}T_s \leq  u_{i}(h+1,t_{k})-u_{i}(h,t_{k}) \leq w^{\max}T_s, \\
\hspace{40mm} h = 0,\ldots,N_{p}-1,  \label{eqDmpc-4}
\end{array}
\end{equation}
\begin{equation} \label{eqDmpc-5}
\begin{array}{ll}
p_{i-1}(h,t_{k})-p_{i}(h,t_{k})-L_{i-1} \geq sm_{0}, \\
\hspace{40mm} h = 1,\ldots,N_{p}+1,
\end{array}
\end{equation}
\begin{equation} \label{eqDmpc-6}
\begin{aligned}
p_{i-1}&(h,t_{k})-p_{i}(h,t_{k})-L_{i-1} \geq sm_{0}+\\
&+\left(\dfrac{v_{i}^2(h,t_{k})}{2b_{i}}+T_{r}v_{i}(h,t_{k})-\dfrac{v_{i-1}^2(h,t_{k})}{2b_{i-1}}\right), \\
&\hspace{4cm} h = 1,\ldots,N_{p}+1,
\end{aligned}
\end{equation}

The initial state constraint at each instant $t_k$ is given as:
\begin{equation}
p_{i}(0,t_{k}) = p_{0,i}, \quad v_{i}(0,t_{k})= v_{0,i}, \label{eqDmpc-7}
\end{equation}
where $p_{0,i}$ and $v_{0,i}$ are the initial position and speed of train $i$ measured at sampling instant $t_k$, respectively.

From \eqref{eq3-8} and \eqref{eq3-9}, the finite-horizon objective function for train $i$ at sampling instant $t_{k}$ can be expressed as
\begin{equation}
J_{i}^{B}(t_{k}) = \sum\nolimits_{h=0}^{N_{p}} \theta_{i}^{B}(h,t_{k}), \label{eqDmpc-8}
\end{equation}
where $\theta^{B}_{i}(h,t_{k})$ is the stage cost at prediction step $h$:
\begin{equation} \nonumber
\theta_{i}^{B}(h,t_{k}) =
\begin{cases}
\zeta_{p,i}\big(p_{i}(h+1,t_{k}) - p_{r}(h+1,t_{k})\big)^{2}+  \\
\hspace{3mm}\zeta_{v,i}\big(v_{i}(h+1,t_{k})-v_{r}(h+1,t_{k})\big)^{2} + \\
\hspace{3mm}\zeta_{u,i}u_{i}^{2}(h,t_{k}),\qquad i=1 \\[2mm]
\zeta_{p,i}\big(p_{i-1}(h+1,t_{k}) - p_{i}(h+1,t_{k})-l \big)^{2}+  \\
\hspace{3mm}\zeta_{v,i}\big(v_{i-1}(h,t_{k})-v_{i}(h,t_{k})\big)^{2}+\\
\hspace{3mm}\zeta_{u,i}u_{i}^{2}(h,t_{k}),\qquad i=2,\ldots,I \\
\end{cases}
\end{equation}

Thus, at each sampling instant $t_{k}$, the discrete-time NMPC problem for multiple virtually coupled trains within the predictive horizon is summarized by the following finite-dimensional nonlinear optimization problem:
\begin{equation}
\begin{aligned}
&\hspace{20mm}\min \; \sum\nolimits_{i=1}^{I}J_{i}^{B}(t_{k}), \\
\text{s.t.}\;
&\begin{cases}
 \text{constraints}~\eqref{eqDmpc-1}-\eqref{eqDmpc-4}, \eqref{eqDmpc-7}, ~i = 1,\ldots,I,\\
\text{constraints}~\eqref{eqDmpc-5}-\eqref{eqDmpc-6},~ i = 2,\ldots,I.
\end{cases}
\end{aligned}
\label{eqDmpc-9}
\end{equation}


The NLP problem~\eqref{eqDmpc-9} is implemented in CasADi and solved with IPOPT at each sampling instant~\citep{Andersson:19}. After solving, only the first optimal control input is applied to each train,
\begin{equation}
u_i(t)=u_i^*(0,t_k), \qquad t\in[t_k,t_{k+1}],
\end{equation}
and the optimization is repeated at the next sampling instant using updated state measurements.



\section{Koopman NMPC Approach}


Based on the continuous-time dynamical system \eqref{eq3-1}, let $x(t) = \big[(p_i(t) )_{i=1}^I, $ $(v_i(t))_{i=1}^I\big]^\top \in \mathbb{R}^{2I}$ denote the state vector, $u(t)  = [u_i(t)]_{i=1}^I \in \mathbb{R}^{I}$ the control input vector, and $r(t)  = [r_i(t)]_{i=1}^I \in \mathbb{R}^I$ the vector of exogenous inputs.

\subsubsection*{Nonlinear State Constraints and Initial Set of Observables:}

The nonlinear state constraints introduce $I$ nonlinear terms -- the squared velocities $(v_i(t)^2 )_{i=1}^I$. The initial set of Koopman observables is therefore given by:
%
    \begin{align}
        &\Phi_\mathrm{in} = \{(p_i)_{i=1}^I, (v_i)_{i=1}^I, (v_i^2)_{i=1}^I\}, ~|\Phi_\mathrm{in}| = 3I, \nonumber \\
        &\phi_\mathrm{in}(x) = [x^\top, \; (v_i(x)^2)_{i=1}^I ]^\top. \nonumber
    \end{align}
Here, with slight abuse of notation, we write $v_i(x)$ for the $(I+i)$-th component of $x$, so that $v_i(x(t))=v_i(t)$.

\subsubsection*{Observables Generation and Lifted System:}

Following the analytical lifting procedure proposed by~\cite{Calogero:25}, an infinite-dimensional basis is defined as:
\begin{align}
        &\Phi = \Phi_\mathrm{in} \cup \{v_i^n\}_{1 \leq i \leq I, \; n \geq 3}, ~|\Phi| = \infty, \nonumber \\
        &\phi(x) = [\phi_\mathrm{in}(x)^\top,\; (v_i(x)^n)_{1\leq i\leq I,\; n\geq 3}]^\top .
\end{align}
The newly generated observables $\{v_i^n\}_{n \geq 3}$ form a complete monomial basis for each $v_i$, $i = 1,\ldots,I$. Based on the complete basis $\Phi$, an infinite-dimensional bilinear Koopman lifted system with state $z(t)=\phi(x(t))$ is derived as:
%
%
\begin{align} \nonumber 
    \dot{z}&(t) = Az(t) + \mathcal{B}(z(t),u(t),r(t)), \\
    \mathcal{B}&(z,u,r):=B_0 \begin{bmatrix}
        u \\ r
    \end{bmatrix} + \sum\nolimits_{i=1}^I B_i z \begin{bmatrix}
        u_i \\ r_i
    \end{bmatrix}. \nonumber
\end{align}

To enhance the practical applicability, dimensionality reduction is adopted. Specifically, with a maximum polynomial degree $\overline{n} \geq 3$, a reduced basis is constructed as:
\begin{align}
        &\Phi = \Phi_\mathrm{in} \cup \{v_i^n\}_{1 \leq i \leq I, \; 3 \leq n \leq \overline{n}}, ~ |\Phi| = (\overline{n}+1)I, \nonumber \\
        &\phi(x) = [\phi_\mathrm{in}(x)^\top, \; (v_i(x)^n)_{1 \leq i \leq I, \; 3 \leq n \leq \overline{n}}]^\top. \nonumber
\end{align}
The reduced Koopman lifted system can be formulated as:
\begin{align} \label{eq:ko_lift_sys_red}
    \dot{z}&(t) = \begin{bmatrix}\dot{z}'(t) \\ \dot{z}''(t) \end{bmatrix} = Az(t) + \nonumber \\
    &+ \underset{\mathcal{B}(z(t),u(t),r(t))}{\underbrace{\begin{bmatrix}
        \mathcal{B}'(z(t),u(t),r(t)) \\ \mathcal{B}''(z(t),u(t),r(t))
    \end{bmatrix}}}  + \begin{bmatrix}
        \bm{0}_{n_z'} \\ f_\mathrm{res}(z(t),u(t),r(t))
    \end{bmatrix}.
\end{align}
Here $f_{\rm res}$ stands for the nonlinear residual terms.

The reduced lifted system~\eqref{eq:ko_lift_sys_red} can be expressed in affine parameter-varying (APV) form, with parameter $(\overline{z}(t),\overline{u}(t),\overline{r}(t))$, by computing its first-order Taylor expansion around $(\overline{z}(t),\overline{u}(t),\overline{r}(t))$, yielding
\begin{align} \label{eq:ko_lift_sys_red_ct_lpv}
    \dot{z}(t) &= \tilde{A}(\overline{z}(t),\overline{u}(t),\overline{r}(t))z(t) + \\
    &\tilde{B}(\overline{z}(t),\overline{u}(t),\overline{r}(t))u(t) + \tilde{b}(\overline{z}(t),\overline{u}(t),\overline{r}(t)). \nonumber
\end{align}
An exact discretization of the APV lifted system~\eqref{eq:ko_lift_sys_red_ct_lpv} can be obtained by computing the closed-form solution of the following convolution integral:
\begin{align} \label{eq:ko_lift_sys_red_dt_lpv}
    & z(t_{k+1}) = e^{\tilde{A}(\overline{z}(t_{k}),\overline{u}(t_{k}),\overline{r}(t_{k}))T_s} z(t_{k}) + \int_{kT_s}^{(k+1)T_s}
    \nonumber \\
    & ~~~~~ e^{\tilde{A}(\overline{z}(t_{k}),\overline{u}(t_{k}),\overline{r}(t_{k}))((k+1)T_s - t)}\big(\tilde{B}(\overline{z}(t_{k}),\overline{u}(t_{k}),\overline{r}(t_{k}))u(t) \nonumber \\
    &~~  + \tilde{b}(\overline{z}(t_{k}),\overline{u}(t_{k}),\overline{r}(t_{k}))\big)dt, \nonumber \\
    & = A_d(\overline{z}(t_{k}),\overline{u}(t_{k}),\overline{r}(t_{k})) z(t_{k}) + B_d(\overline{z}(t_{k}),\overline{u}(t_{k}),\overline{r}(t_{k}))u(t_{k}) \nonumber  \\
    &~~ + b_d(\overline{z}(t_{k}),\overline{u}(t_{k}),\overline{r}(t_{k})).
\end{align}
%

\subsubsection*{K-NMPC Formulation:} The lifted state is split as:
\begin{align}
&z(h,t_{k}) = [z_p^\top(h,t_{k}) , \; z_v^\top(h,t_{k}) , \; z_\sigma^\top(h,t_{k}), \; z_\psi^\top(h,t_{k}) ]^\top, \nonumber \\
&z_x(h,t_{k})  = [z_p^\top(h,t_{k}) , \; z_v^\top(h,t_{k}) ]^\top, \nonumber
\end{align}
where
    \begin{align}
       &z_p(h,t_{k}) = [p_i(h,t_{k})]_{i=1}^I, ~~~z_v(h,t_{k})  = [v_i(h,t_{k}) ]_{i=1}^I, \nonumber \\
        &z_\sigma(h,t_{k})  = [v_i(h,t_{k})^2]_{i=1}^I, ~z_\psi(h,t_{k}) = [v_i^n(h,t_{k})]_{\substack{1 \leq i \leq I, \\ 3 \leq n \leq \overline{n}}}. \nonumber
    \end{align}

Then, we can formulate, at each sampling time instant $t_{k}$, the reduced K-NMPC optimal control problem, as follows:
\begin{subequations}\label{eq:knmpc_red}
    \begin{align}
        & \hspace{20mm} \min_{z,u} \sum\nolimits_{i=1}^{I}J_{i}^{K}(t_{k}),  \tag{\ref{eq:knmpc_red}} \\
        \textrm{s.t.}
        & z(h+1,t_{k}) = \bar{A}(h,t_{k}) z(h,t_{k})  + \nonumber \\
        & \hspace{5mm} \bar{B}(h,t_{k}) u(h,t_{k}) + \bar{b}(h,t_{k}), ~h=0,\ldots,N_p, \label{eq:knmpc_red_c} \tag{\ref{eq:knmpc_red}a} \\
        & z(0,t_{k}) = \phi(x(t_{k})), \label{eq:knmpc_red_b} \tag{\ref{eq:knmpc_red}b}   \\
        & 0 \leq z_{v,i}(h,t_{k}) \leq v_{i}^{max}(h,t_{k}),\nonumber \\
        & \hspace{20mm} i=1,\ldots,I, h=1,\ldots,N_p+1, \tag{\ref{eq:knmpc_red}c}\\
        & u_{i}^{\min} \leq u_{i}(h,t_{k}) \leq u_{i}^{\max}, \nonumber \\
        & \hspace{20mm} i=1,\ldots,I, h=0,\ldots,N_p,\tag{\ref{eq:knmpc_red}d} \\
        & w^{\min}T_s \leq u_{i}(h+1,t_{k}) - u_{i}(h,t_{k}) \leq w^{\max}T_s, \nonumber \\
        & \hspace{20mm} i=1,\ldots,I, h=0,\ldots,N_p-1, \tag{\ref{eq:knmpc_red}e} \\
        & z_{p,i-1}(h,t_{k}) - z_{p,i}(h,t_{k}) - L \geq sm_0, \nonumber \\
        & \hspace{20mm} i=2,\ldots,I, h=1,\ldots,N_p+1, \tag{\ref{eq:knmpc_red}f}  \\
        & z_{p,i-1}(h,t_{k}) - z_{p,i}(h,t_{k}) - L \geq sm_0 + \frac{z_{\sigma,i}(h,t_{k})}{2b_{i}}  \nonumber \\
        & \hspace{15mm}+ T_r z_{v,i}(h,t_{k})- \frac{z_{\sigma,i-1}(h,t_{k})}{2b_{i-1}},  \nonumber \\
        & \hspace{20mm} i=2,\ldots,I, h=1,\ldots,N_p+1, \tag{\ref{eq:knmpc_red}g}
    \end{align}
\end{subequations}
where $J_{i}^{K}(t_{k}) = \sum_{h=0}^{N_{p}} \theta _{i}^{K}(h,t_{k})$. The stage cost $\theta _{i}^{K}(h,t_{k})$ for the K-NMPC optimal control problem of each virtually coupled train $i$ is given as:
\begin{equation} \nonumber
\theta_{i}^{K}(h,t_{k}) =
\begin{cases}
\zeta_{p,i}\big( z_{p,i}(h+1,t_{k}) -  p_{r}(h+1,t_{k}) \big)^{2}+  \\
\hspace{3mm}\zeta_{v,i}\big( z_{v,i}(h+1,t_{k}) -v_{r}(h+1,t_{k}) \big)^{2} + \\
\hspace{3mm}\zeta_{u,i}u_{i}^{2}(h,t_{k})\qquad i=1, \\[2mm]
\zeta_{p,i}\big( z_{p,i-1}(h+1,t_{k}) - z_{p,i}(h+1,t_{k}) -l  \big)^{2}+  \\
\hspace{3mm}\zeta_{v,i}\big(z_{v,i-1}(h+1,t_{k}) - z_{v,i}(h+1,t_{k}) \big)^{2}+\\
\hspace{3mm}\zeta_{u,i} u_{i}^{2}(h,t_{k}),\qquad i = 2,\ldots,I.
\end{cases}
\end{equation}

The prediction model~\eqref{eq:knmpc_red_c} is obtained from~\eqref{eq:ko_lift_sys_red_dt_lpv}:
\begin{subequations} \nonumber
    \begin{align}
        \bar{A}(h,t_{k}) &= A_d(\overline{z}(t_{k}),\overline{u}(t_{k}),\overline{r}(t_{k})),\\
        \bar{B}(h,t_{k}) &= B_d(\overline{z}(t_{k}),\overline{u}(t_{k}),\overline{r}(t_{k})),\\
        \bar{b}(h,t_{k}) &= b_d(\overline{z}(t_{k}),\overline{u}(t_{k}),\overline{r}(t_{k})).
    \end{align}
\end{subequations}
The trajectories $(\overline{z}(h,t_{k}),\overline{u}(h,t_{k}))_{h=0}^{N_p-1}$ are obtained from the shifted optimal trajectory computed at the previous sampling instant $t_{k-1}$ combined with the current state measurements, and $\overline{r}(t_{k})$ is the current value of the exogenous inputs.


\section{Simulation experiments}



\begin{figure}[h!]
\centering
\includegraphics[width=0.8\linewidth]{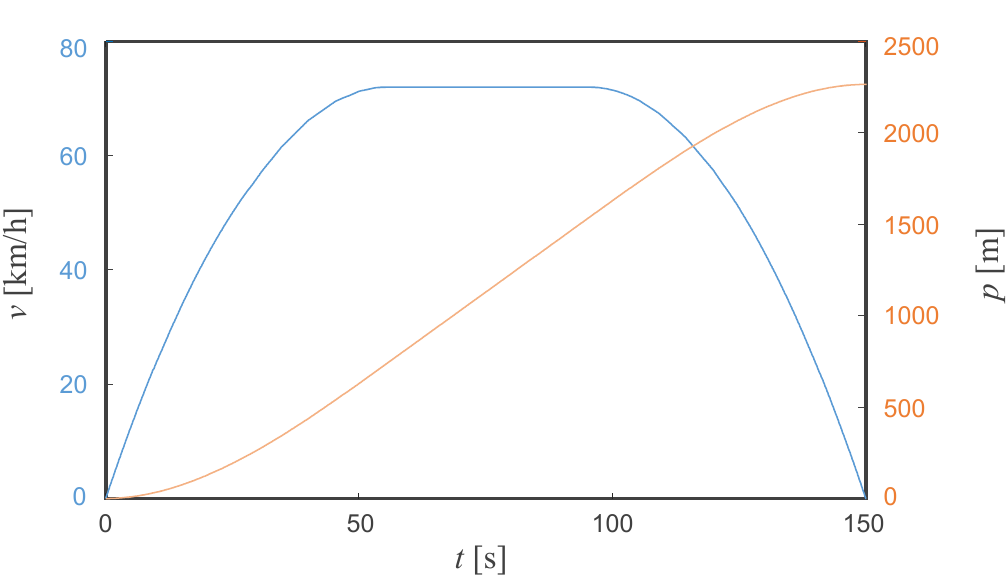}
\caption{\label{fig:ref} Reference trajectory of the leading train.}
\end{figure}

Realistic operating data from the Beijing Metro Yizhuang Line are employed for the section between Tongji Nan Station and Jinhai Lu Station of length of 2265 m. The reference trajectory (Fig.~\ref{fig:ref}) for the leading train was planned in the MATLAB package GPOPS,
implementing the Radau pseudospectral optimal control method~\citep{YeHB:16}, with a prescribed travel time of 150s. Other parameters are summarized in Table~\ref{tab:sim_params}. All simulations are performed in MATLAB R2020a on a computer with 16 GB of RAM and a 2.4 GHz processor.

\begin{figure}[t!]
\centering
\begin{subfigure}{0.95\linewidth}
    \centering
    \includegraphics[width=\linewidth]{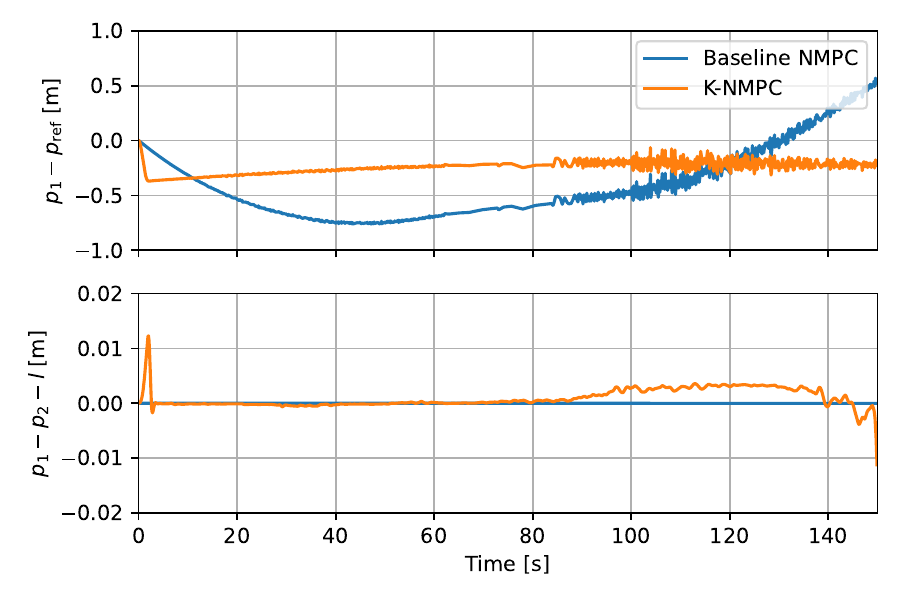}
    \caption{Spacing deviations.}
    \label{fig:sdev}
\end{subfigure}
\begin{subfigure}{0.95\linewidth}
    \centering
    \includegraphics[width=\linewidth]{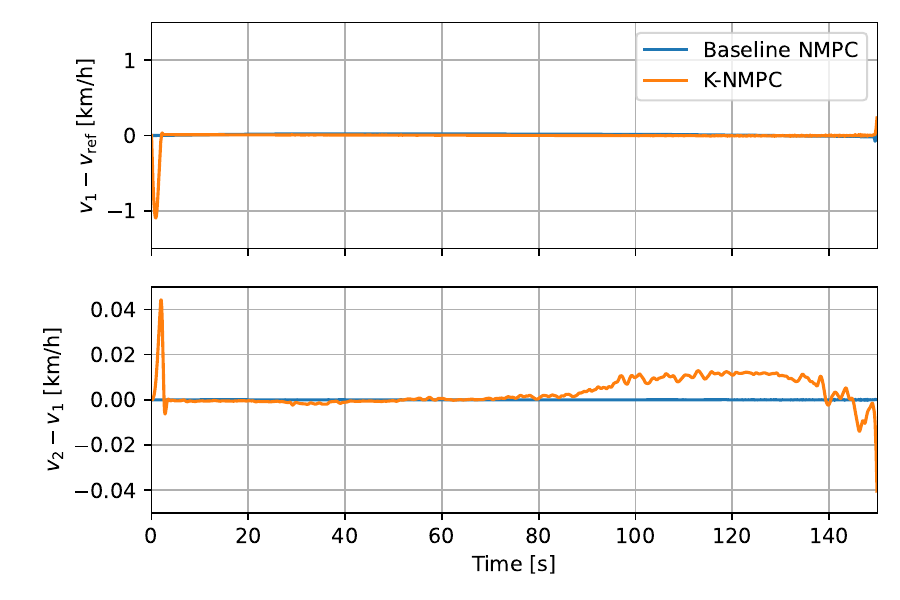}
    \caption{Speed deviations.}
    \label{fig:vdev}
\end{subfigure}
\caption{Tracking performance comparison}
\label{fig:tracking}
\end{figure}
\begin{figure}[t!]
\centering
\begin{subfigure}{0.95\linewidth}
    \centering
    \includegraphics[width=\linewidth]{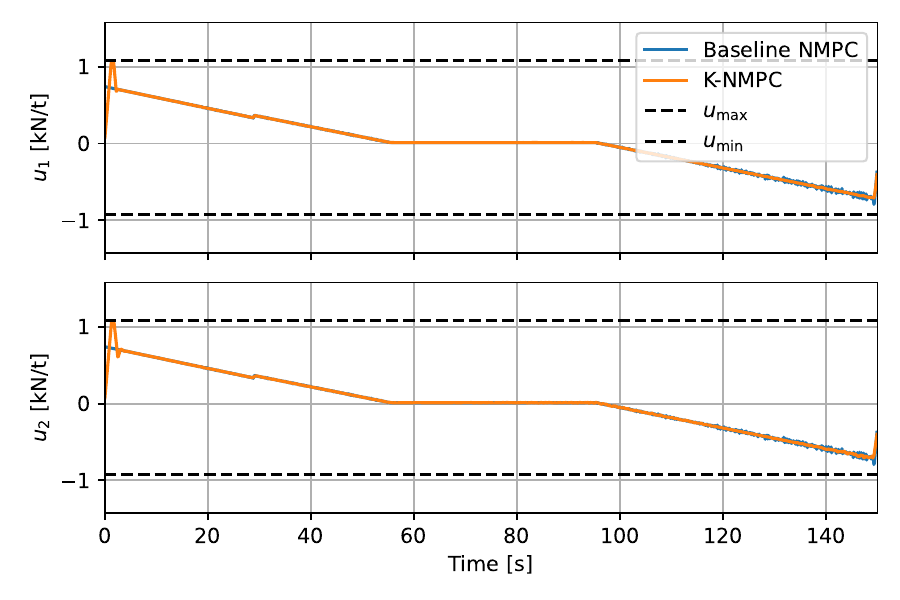}
    \caption{Control forces.}
    \label{fig:control_force}
\end{subfigure}
\begin{subfigure}{0.95\linewidth}
    \centering
    \includegraphics[width=\linewidth]{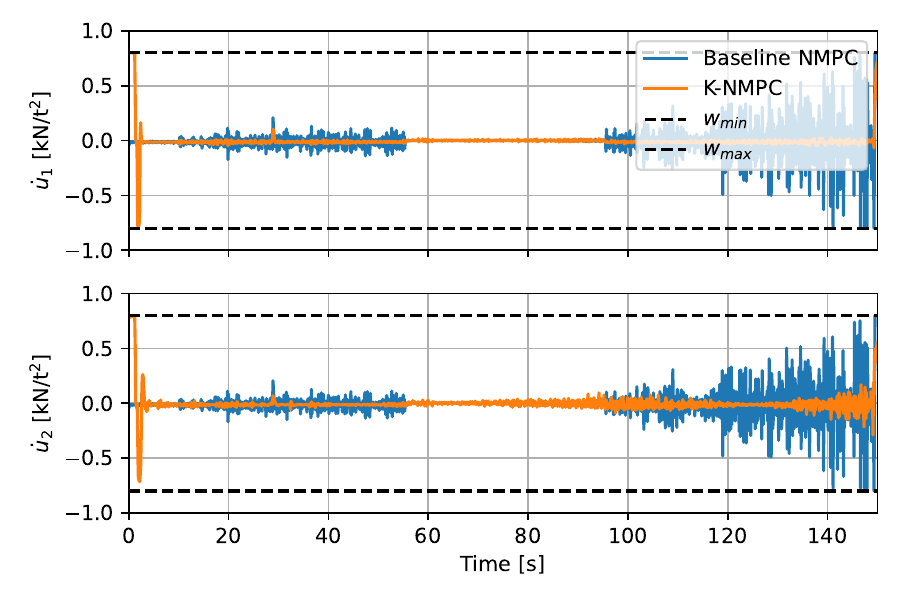}
    \caption{Rates of change of control forces.}
    \label{fig:control_rate}
\end{subfigure}

\caption{Control inputs and their rates of change}
\label{fig:control}
\end{figure}

Fig.~\ref{fig:tracking} compares the spacing and speed deviations. K-NMPC yields a smaller leading-train position deviation $p_1-p_{\mathrm{ref}}$, while the following-train spacing error $p_1-p_2-l$ remains close to zero for both controllers. The speed deviations are also small under both methods; K-NMPC shows a larger initial transient in $v_1-v_{\mathrm{ref}}$, but the deviation quickly decays, and the relative speed error $v_2-v_1$ remains below 0.05 km/h. Fig.~\ref{fig:control} shows that both controllers satisfy the force and rate constraints. K-NMPC produces larger initial variations, but smaller control variations over most of the trip, indicating smoother operation after the initial transient.
Table~\ref{tab:tracking_err} reports the average absolute spacing and speed deviations and the energy-related index $\int_{t_0}^{t_f}u^2(t)dt$. K-NMPC reduces the spacing deviation by about 49\% with nearly identical speed deviation and energy index.
\begin{table}[htb]
  \centering
  \caption{Comparison of tracking performance.}
  \label{tab:tracking_err}
  \begin{tabular}{cccccc}
    \toprule
    Method
      & $ \|dev_{p}\|_{1}$  [m]
      & $ \|dev_{v}\|_{1}$ [km/h]
      & $ \int_{t_{0}}^{t_{f}} u^{2}(t)dt$ \\
    \midrule
    Baseline NMPC  & 0.237  & 0.008 & 388.84 \\
    K-NMPC         & 0.122  & 0.009 & 389.66 \\
    \bottomrule
  \end{tabular}
\end{table}

\begin{figure}[h]
\centering
\includegraphics[width=0.8\linewidth]{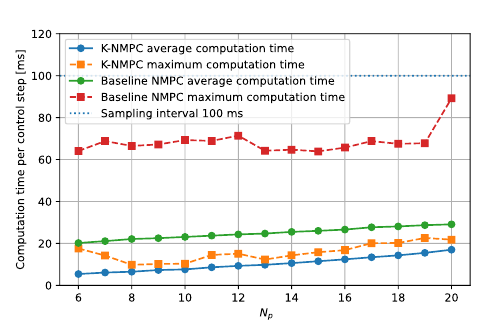}
\caption{\label{fig:time} Computation time for different predictive horizons}
\end{figure}

Fig.~\ref{fig:time} compares the computation times for prediction horizons $N_p=6,\ldots,20$. K-NMPC scales more favorably: its average computation time increases from 5.4 to 17.0 ms, while the maximum remains below 23 ms. The baseline NMPC requires 20.2--29.1 ms on average, with maximum times of about 64--89 ms; the K-NMPC reduces the average computation time by 40\%--70\% and the maximum time by up to 85\%, supporting real-time implementation.

\section{Conclusion and Discussion}

\begin{table*}
\centering
\begin{tabular}{llll}
\toprule
Parameter & Value & Parameter & Value \\
\midrule
$I$ & $2$ & Line length & $2265~\mathrm{m}$ \\
Trip time & $150~\mathrm{s}$ & $T_s$ & $0.1~\mathrm{s}$ \\
$N_p$ & $10$ & $\bar n$ & $3$ \\
$L$ & $18.0~\mathrm{m}$ & $sm_0$ & $4~\mathrm{m}$ \\
$d_{\mathrm{des}}$ & $9~\mathrm{m}$ & $b$ & $1.0~\mathrm{m/s^2}$ \\
$T_r$ & $0.2~\mathrm{s}$ & $[u_{\min},u_{\max}]$ & $[-1.10,0.93]~\mathrm{kN/t}$ \\
$[w_{\min},w_{\max}]$ & $[-0.8,0.8]~\mathrm{kN/(t\,s)}$ & $(c_0,c_1,c_2)$ & $(1.9904{\times}10^{-2},\,2.1944{\times}10^{-3},\,2.2950{\times}10^{-4})$ \\
\bottomrule
\end{tabular}
\caption{Main simulation parameters.}
\label{tab:sim_params}
\end{table*}

This paper proposed an analytical Koopman-based NMPC scheme for virtually coupled train control. By lifting the nonlinear train dynamics and velocity-square safety terms into a reduced Koopman observable space, the proposed method converts the online prediction problem into a quadratic program after freezing the APV model along the shifted predicted trajectory. Simulation results for a two-train metro scenario show that K-NMPC achieves tracking performance comparable to baseline NMPC while substantially reducing computation time.

\bibliography{ifacconf}

\begin{thebibliography}{25}
\providecommand{\natexlab}[1]{#1}
\providecommand{\url}[1]{\texttt{#1}}
\providecommand{\urlprefix}{URL }
\expandafter\ifx\csname urlstyle\endcsname\relax
  \providecommand{\doi}[1]{doi:\discretionary{}{}{}#1}\else
  \providecommand{\doi}{doi:\discretionary{}{}{}\begingroup
  \urlstyle{rm}\Url}\fi

\bibitem[{Andersson et~al.(2019)Andersson, Gillis, Horn, Rawlings, and
  Diehl}]{Andersson:19}
Andersson, J., Gillis, J., Horn, G., Rawlings, J., and Diehl, M. (2019).
\newblock Cas{AD}i: A software framework for nonlinear optimization and optimal
  control.
\newblock \emph{Mathematical Programming Computation}, 11(1), 1--36.

\bibitem[{Boggio et~al.(2023)Boggio, Novara, and Taragna}]{Boggio:23}
Boggio, M., Novara, C., and Taragna, M. (2023).
\newblock Trajectory planning and control for autonomous vehicles: a “fast”
  data-aided {NMPC} approach.
\newblock \emph{European Journal of Control}, 74, 100857.

\bibitem[{Calogero et~al.(2025)Calogero, Boggio, Novara, and
  Rizzo}]{Calogero:25}
Calogero, L., Boggio, M., Novara, C., and Rizzo, A. (2025).
\newblock A general analytical framework for fast solving nonlinear {MPC}
  problems in the linear {K}oopman space.
\newblock In \emph{Proceedings of the European Control Conference (ECC)}.

\bibitem[{Faria et~al.(2024)Faria, Capron, Secchi, and Jr.}]{Faria:24}
Faria, R., Capron, B., Secchi, A., and Jr., M.D.S. (2024).
\newblock A data-driven tracking control framework using physics-informed
  neural networks and deep reinforcement learning for dynamical systems.
\newblock \emph{Engineering Applications of Artificial Intelligence}, 127,
  107256.

\bibitem[{Felez et~al.(2019)Felez, Kim, and Borrelli}]{Felez:19}
Felez, J., Kim, Y., and Borrelli, F. (2019).
\newblock A model predictive control approach for virtual coupling in railways.
\newblock \emph{IEEE Trans. on Intelligent Transp. Syst.}, 20, 2728--2739.

\bibitem[{Gao et~al.(2023)Gao, Peng, Peng, Chen, and Huang}]{GaoDZ:23}
Gao, D., Peng, J., Peng, H., Chen, B., and Huang, Z. (2023).
\newblock Cooperative braking of urban rail vehicles with {K}oopman model
  predictive control.
\newblock \emph{IET Control Theory \& Applications}, 17(15), 2005--2016.

\bibitem[{Gao et~al.(2025)Gao, Jia, Li, Yang, and Xie}]{Gao:25}
Gao, L., Jia, B., Li, D., Yang, Y., and Xie, S. (2025).
\newblock Efficient safety-critical trajectory planning for any n-trailer
  system with a general model.
\newblock \emph{Control Engineering Practice}, 158, 106287.

\bibitem[{Goikoetxea(2016)}]{Goikoetxea:16}
Goikoetxea, J. (2016).
\newblock Roadmap towards the wireless virtual coupling of trains.
\newblock In J.~Mendizabal, M.~Berbineau, A.~Vinel, S.~Pfletschinger,
  H.~Bonneville, A.~Pirovano, S.~Plass, R.~Scopigno, and H.~Aniss (eds.),
  \emph{Roadmap towards the wireless virtual coupling of train}, 3--9. Springer
  International Publishing, Cham, 3rd edition.

\bibitem[{Goswami and Paley(2021)}]{paper-goswami-2021}
Goswami, D. and Paley, D.A. (2021).
\newblock {Bilinearization, Reachability, and Optimal Control of Control-Affine
  Nonlinear Systems: A {K}oopman Spectral Approach}.
\newblock \emph{{IEEE Trans. Autom. Control}}, 67(6), 2715--2728.

\bibitem[{Iwnicki et~al.(2019)Iwnicki, Spiryagin, Cole, and
  McSweeney}]{Iwnicki:19}
Iwnicki, S., Spiryagin, M., Cole, C., and McSweeney, T. (2019).
\newblock \emph{Handbook of Railway Vehicle Dynamics}.
\newblock CRC Press, Boca Raton, 2nd edition.

\bibitem[{Korda and Mezi{\'c}(2018)}]{Korda:18}
Korda, M. and Mezi{\'c}, I. (2018).
\newblock Linear predictors for nonlinear dynamical systems: Koopman operator
  meets model predictive control.
\newblock \emph{Automatica}, 93, 149--160.

\bibitem[{Kuutti et~al.(2021)Kuutti, Bowden, Jin, Barber, and
  Fallah}]{Kuutti:21}
Kuutti, S., Bowden, R., Jin, Y., Barber, P., and Fallah, S. (2021).
\newblock A survey of deep learning applications to autonomous vehicle control.
\newblock \emph{IEEE Trans. on Intelligent Transp. Syst.}, 22(2), 712--733.

\bibitem[{Li et~al.(2015)Li, Yang, and Gao}]{LiSK:15}
Li, S., Yang, L., and Gao, Z. (2015).
\newblock Coordinated cruise control for high-speed train movements based on a
  multi-agent model.
\newblock \emph{Transp. Research Part C: Emerging Technologies}, 56, 281--292.

\bibitem[{Li et~al.(2020)Li, Yang, and Gao}]{LiSK:20}
Li, S., Yang, L., and Gao, Z. (2020).
\newblock Distributed optimal control for multiple high-speed train movement:
  An alternating direction method of multipliers.
\newblock \emph{Automatica}, 112, 108646.

\bibitem[{Liu et~al.(2022)Liu, Liu, Su, Wei, and Tang}]{LiuYF:22}
Liu, Y., Liu, R., Su, S., Wei, C., and Tang, T. (2022).
\newblock Distributed model predictive control strategy for constrained
  high-speed virtually coupled train set.
\newblock \emph{IEEE Trans. on Vehicular Technology}, 71, 171--183.

\bibitem[{Mitchell et~al.(2016)Mitchell, Goddard, Montes, Stanley, Muttram,
  Coenraad, Por{\'e}, Andrews, and Lochman}]{Mitchel:16}
Mitchell, I., Goddard, E., Montes, F., Stanley, P., Muttram, R., Coenraad, W.,
  Por{\'e}, J., Andrews, S., and Lochman, L. (2016).
\newblock Ertms level 4, train convoys or virtual coupling.
\newblock \emph{IRSE NEWS}, 219, 1--3.

\bibitem[{Park et~al.(2023)Park, Lee, and Eun}]{Park:22}
Park, J., Lee, B., and Eun, Y. (2023).
\newblock Virtual coupling of railway vehicles: Gap reference for merge and
  separation, robust control, and position measurement.
\newblock \emph{IEEE Trans. on Intelligent Transp. Syst.}, 2, 1085--1096.

\bibitem[{Plissonneau et~al.(2024)Plissonneau, Jourdan, Trentesaux, Abdi,
  Sallak, Bekrar, Quost, and Sch{\"o}n}]{Pliss:24}
Plissonneau, A., Jourdan, L., Trentesaux, D., Abdi, L., Sallak, M., Bekrar, A.,
  Quost, B., and Sch{\"o}n, W. (2024).
\newblock Deep reinforcement learning with predictive auxiliary task for
  autonomous train collision avoidance.
\newblock \emph{Journal of Rail Transp. Planning \& Management}, 31, 100453.

\bibitem[{Santoso et~al.(2018)Santoso, Garratt, and Anavatti}]{Santoso:18}
Santoso, F., Garratt, M., and Anavatti, S. (2018).
\newblock State‐of‐the‐art intelligent flight control systems in unmanned
  aerial vehicles.
\newblock \emph{IEEE Trans. on Automation Science and Engineering}, 15(2),
  613--627.

\bibitem[{Su et~al.(2022)Su, Liu, Zhu, Li, Tang, and Lv}]{Su:22}
Su, S., Liu, W., Zhu, Q., Li, R., Tang, T., and Lv, J. (2022).
\newblock An adaptive safety control approach for virtual coupling system with
  model parametric uncertainties.
\newblock \emph{Accident Analysis \& Prevention}, 173, 106703.

\bibitem[{Wang et~al.(2025)Wang, Wang, Li, Qu, Xu, Hu, and Yang}]{WangP:25}
Wang, P., Wang, J., Li, Z., Qu, T., Xu, F., Hu, Y., and Yang, H. (2025).
\newblock Koopman-{MPC}-based energy-efficient integrated control of attitude
  maneuver and vibration suppression for nonlinear in-wheel motor-active
  suspension on uneven roads.
\newblock \emph{Energy}, 336, 138157.

\bibitem[{Ye and Liu(2016)}]{YeHB:16}
Ye, H. and Liu, R. (2016).
\newblock A multiphase optimal control method for multi-train control and
  scheduling on railway lines.
\newblock \emph{Transp. Research Part B: Methodological}, 93, 377--393.

\bibitem[{Zhang et~al.(2023)Zhang, Wang, Zhang, and Chai}]{ZhangQH:23}
Zhang, Q., Wang, H., Zhang, Y., and Chai, M. (2023).
\newblock An adaptive safety control approach for virtual coupling system with
  model parametric uncertainties.
\newblock \emph{Transp. Research Part C: Emerging Technologies}, 154, 104235.

\bibitem[{Zhang et~al.(2025)Zhang, Wang, Du, and Zheng}]{ZhangW:25}
Zhang, W., Wang, Q., Du, X., and Zheng, Y. (2025).
\newblock Real-time {NMPC} for three-dimensional trajectory tracking control of
  {AUV} with disturbances.
\newblock \emph{Ocean Engineering}, 319, 120267.

\bibitem[{Zheng et~al.(2024)Zheng, Li, Zheng, and Hashemi}]{ZhengH:24}
Zheng, H., Li, Y., Zheng, L., and Hashemi, E. (2024).
\newblock Koopman‐based hybrid modeling and zonotopic tube robust {MPC} for
  motion control of automated vehicles.
\newblock \emph{IEEE Trans. on Intelligent Transp. Syst.}, 25(10), 13598.

\end{thebibliography}

\end{document}